# Torsional oscillator and synchrotron x-ray experiments on solid $^4$He in aerogel

N. Mulders[1], J.T. West[2], M.H.W. Chan[2], C.N. Kodituwakku[3], C.A. Burns[3], and L.B. Lurio[4]

[1]*Dept. of Physics and Astronomy, University of Delaware, Newark, Delaware 19716*

[2]*Dept. of Physics, The Pennsylvania State University, University Park, Pennsylvania 16802*

[3]*Dept. of Physics, Western Michigan University, Kalamazoo, Michigan 49008*

[4]*Dept. of Physics, Northern Illinois University, Dekalb, Illinois 60615*

X-ray diffraction experiments show that solid $^4$He grown in aerogel is highly polycrystalline, with a hcp crystal structure (as in bulk) and a crystallite size of approximately 100 nm. In contrast to the expectation that the highly disordered solid will have a large supersolid fraction, torsional oscillator measurements show a behavior that is strikingly similar to high purity crystals grown from the superfluid phase. The low temperature supersolid fraction is only ~$3\times10^{-4}$ and the onset temperature is ~ 100 mK.

PACS numbers: 67.80.-s, 67.80.M

Over the last four years, a number of groups have reported small frequency increases in torsional oscillator (TO) experiments on solid $^4$He below temperatures of the order of 200 mK[1-7]. The observations are consistent with the decoupling of some of the helium from the motion of the oscillator, and the appearance of a supersolid state. The magnitude of the effect is cell dependent and also strongly dependent on sample preparation and history. Samples presumed to be the most pristine tend to show smaller nonclassical rotational inertia fraction (NCRIf), as small as 0.03%[8], whereas rapidly quenched, and possibly highly disordered samples may have an NCRIf as large as 20%[4]. In one experiment, the NCRIf was dramatically reduced by annealing the sample [3]. These results indicate that defects such as dislocation lines play an important role in the formation of the supersolid state. Indeed, the current theoretical picture is that perfect crystals will not exhibit supersolidity[9, 10], and that the vacancy concentration at low temperatures is exceedingly small[11]. On the other hand, simulations of dislocations[12, 13] and large angle grain boundaries[14] indicate that these types of structures may support supersolidity.

In addition to the TO oscillator results, Day and Beamish recently observed an approximately 10% increase in the shear modulus of $^4$He at low temperatures[15]. The dependence on temperature, $^3$He concentration [16] and stress amplitude [17] is similar to what is observed in the TO experiments. They suggest an explanation in terms of a dislocation network that is pinned by $^3$He, but which becomes mobile when the $^3$He evaporates off the dislocation lines. The mobile dislocations can respond to stress, thus lowering the shear modulus. The suppression of the NCRIf is surmised to be due to an interaction between the now mobile dislocation lines and the superflow. This suggestion is consistent with the results of a TO study on samples with $^3$He concentrations that range from 1 parts per billion up to 30 parts per million. It was found that the onset temperature of NCRI increases with $^3$He impurities and appears to track the condensation of $^3$He atoms onto the dislocation lines.[16] $^3$He as well as dislocation lines may serve as pinning sites of vortices[18]. There is experimental evidence suggesting that the suppression of NCRIf is due to the unpinning of vortices.[5, 17] One thus ends up with a model in which on the one hand defects are necessary for supersolidity, while on the other hand the mobility of at least one class of defects is effective in destroying the supersolid response.

Both aspects are open to experimental verification. An introduction of a high density of stable defects should lead to a large supersolid response. Furthermore, immobilizing the dislocation network should lead to a high transition temperature. In an attempt to realize this situation, we have grown solid $^4$He in a silica aerogel on the assumption that the silica network would sufficiently impede the growth of high quality crystals, while forming effective anchoring points for line dislocations.

One may argue that there is already one experiment that does substantially the same thing, namely the TO experiment on solid $^4$He in Vycor [2]. However, the situation in Vycor may be quite different from that of the bulk solid. Neutron diffraction experiments by Wallacher et al. indicate that in a porous matrix similar to Vycor only a small fraction, ~30%, of the helium is crystalline[19]. This crystalline phase appears to be bcc and is stable down to 60 mK. This phase is expected to have quite different defect structures than the hcp bulk solid. While the similarity between the TO experiments in bulk and in Vycor seem to point towards a common origin, the significant differences between the two systems impel us to put aside the Vycor results for now and focus on a system in which the deviation from the bulk is likely to be more tractable.

Aerogel[20] has played an important role in the investigations of phase transitions in quantum fluids[21, 22]. Consisting of nanometer size silica particles, which self-assemble to form an open structure consisting of interpenetrating fractal clusters with a correlation length of the order of 1000 Å, they provide a source of static or quenched disorder. Because of their low (electron) density, aerogels are substantially transparent to x-rays, which makes it possible to do both structural and thermal/mechanical measurements on the same sample. The TO and x-ray diffraction experiments reported here were done on 95% porous, base catalyzed silica aerogels. The specific surface area of the samples is approximately 1000 m$^2$/g corresponding to a primary SiO$_2$ particle diameter of ~30 Å. The large surface-to-volume ratio, 100 m$^2$/cm$^3$, implies that ~3% of the solid helium is within 3 Å of a silica surface. While aerogels are normally thought of as having a very open structure, numerical simulations show that for our samples no helium is more than ~200 Å away from silica[23, 24]. The mean free path, which in this context we would like to interpret as the mean distance between pinning sites is approximately 500 Å. This may be compared with a typical loop length in a dislocation network of 5 μm [15, 25].

In both the TO and the x-ray experiments cylindrical aerogel samples were grown under identical conditions, with a diameter that fits snuggly inside the cylindrical sample cells. The heights of the aerogel samples are slightly greater than that of the cells. Sealing the cells then compressed the aerogel disks by 5 – 10% securing them rigidly in place. Another similarly grown aerogel sample of the same porosity was placed in a separate cell and a solid helium sample was grown. The cell was then warmed up and the aerogel was visually inspected. It appears that growing solid helium inside aerogel does not damage it. The samples were grown using commercially available ultra-high-purity (UHP) helium gas which has a typical $^3$He concentration of ~0.3 ppm.

The x-ray diffraction experiments were done on beamline 33-ID-E at the Advanced Photon Source at Argonne National Laboratory. Measurements were made at an x-ray energy of 30 keV. Although using a high photon energy decreases the resolution of the experiment, it results in a sharp reduction of the heat load and allows a larger range of Bragg peaks to be accessed through our windows. Depending on the information desired, either a Princeton Instruments CCD camera or a higher resolution setup employing a slit collimated Oxford Cyberstar scintillation point detector was used.

We built a dilution refrigerator specifically designed for x-ray work at the Advanced Photon Source. This cryostat is based on the CryoMech PT405 pulse tube refrigerator. This completely eliminates the need for liquid cryogens. The dilution refrigerator can continuously operate down to 45 mK. The cryostat can be cooled down ahead of the actual experiment, and mounted on to an x-ray spectrometer while remaining cold. Aluminized mylar windows provide x-ray access to the sample space.

The x-ray cell consists of a copper disk with a cylindrical hole, 3 mm in diameter. Two beryllium windows seal the cell, forming a cavity with a thickness of 2 mm. The x-ray beam, 0.1 x 0.1 mm in cross section passes through the center of the cell, and the diffracted photons are detected in transmission. Figure 1 shows the resulting powder pattern from a solid grown at a constant pressure of 60 bar. Two sets of wider rings are due to the beryllium windows (fabricated by a powder technology based process) but within these rings one can observe the first four (partial) rings corresponding to hcp $^4$He, with the fifth (101) reflection just within the second set of Be rings. The rings are incomplete and not uniform but vary in width and intensity. This indicates the existence of some texture such as preferred growth directions for the crystallites. The image was generated by subtracting the diffraction pattern just before freezing from that of the solid sample. This greatly reduces the intensity of the Be rings and brings out the much fainter solid helium ones. The dark region coinciding with the (100) ring is the negative of the first peak in the liquid structure factor.

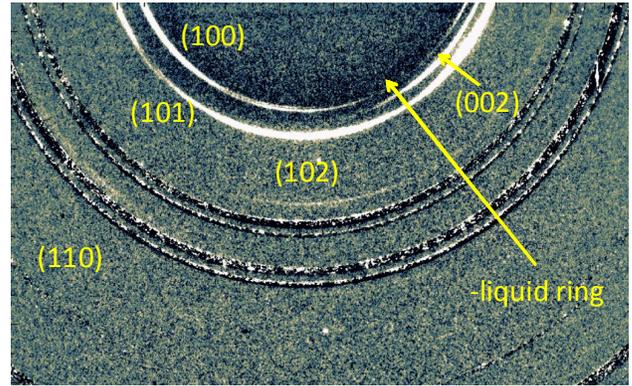

*Figure 1. The powder pattern for a solid helium sample in aerogel at 60 bar. The rings corresponding to the first 5 reflections of a hcp structure are indicated.*

The average crystallite size, *D*, can be determined from the width of the diffraction ring using

$$D = \frac{S\lambda}{W_{FWHM} \cos\theta} \quad (1)$$

where $S$ is a shape factor of order one, $\lambda$ the x-ray wavelength, $W_{FWHM}$ the width of the peak, and $\theta$ the diffraction angle. A typical diffraction peak measured with high resolution is shown in fig. 2 (for the (101) reflection). The diffraction peaks are about five times wider than the instrumental width. Using eq. 1, we find a crystallite size of 1000 Å. Widths determined from other reflections were consistent with this value.

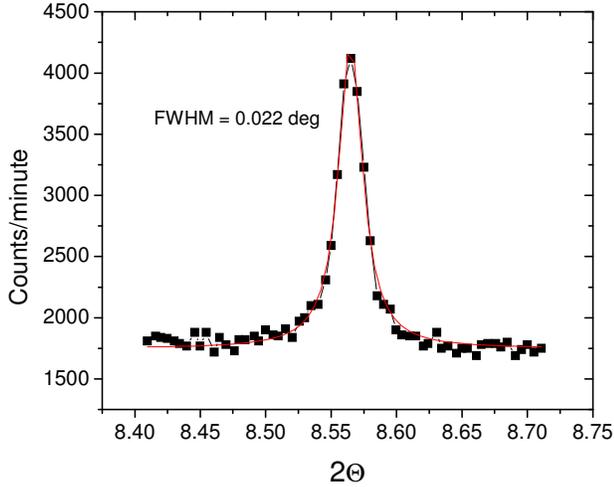

*Figure 2. The diffraction peak for the (101) reflection. The width of the peak indicates a crystallite size of ~1000Å.*

Torsional oscillator experiments were done at pressures from 34 to 48 bar. The data are shown in figure 3, together with the results for bulk solid UHP $^4$He samples grown from both the superfluid phase at constant pressure (CP) (at 25.6 bar), and one grown using the blocked capillary (BC) method (33.7 bar sample). Showing a sharp transition and small NCRIf, the highly disordered solid grown in aerogel resembles a high quality crystal rather than a presumably poor quality blocked capillary one. We observe thermal history dependence and a critical velocity effect qualitatively similar to Clark et al. [17] for drive levels above ~100 μm/s.

The very small value of the NCRIf is all the more surprising when one estimates the fraction of the solid that is either associated with grain boundaries, or adjacent to the silica aerogel network. With a crystallite size of 1000 Å and a width of the grain boundary of the order of 3 Å, it follows that about 1% of the solid resides on a grain boundary, i.e., a much larger fraction than the actually observed NCRIf in these samples. A similar fraction of the solid resides near the silica. Simulations by Khairallah and Ceperley indicate that near the silica a layered structure forms with a disordered, delocalized layer between the crystalline solid and a highly localized first layer. The superfluid fraction in this disordered layer could be as large as 0.3 [26]. However, it is known that the tortuosity of liquid films on aerogels is large, and that effectively only a small fraction of the superfluid decouples from the motion of a TO [27, 28].

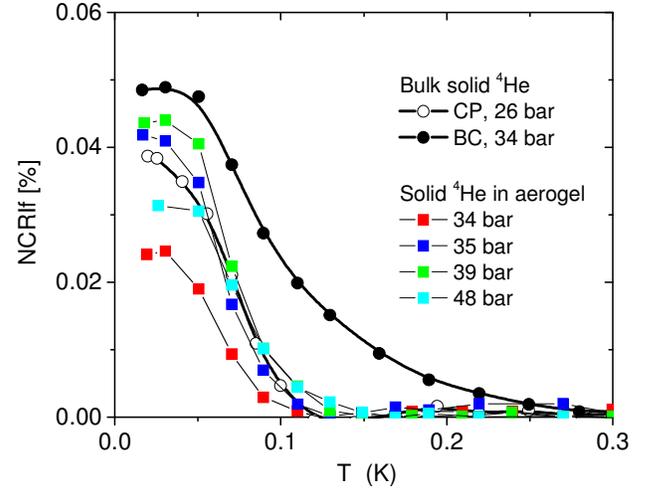

*Figure 3. The NCRIf from solid $^4$He grown in aerogel at pressures between 34 and 48 bar. The bulk solid data is taken from Ref.[8], Fig 2b. To emphasize its different temperature dependence, the BC data **is scaled** down by a factor of 3. All data was taken at low velocity (rim velocity<10 μm/s) BC: blocked capillary, CP: constant pressure.*

The recent TO experiment on samples with different $^3$He concentrations [16] and shear modulus measurements [15] indicate that the dislocation network is the most relevant disorder for the appearance of NCRI. Recent PIMC simulation studies indicate the cores of the dislocation lines support superfluidity.[12] It is reasonable to speculate that the observed NCRI in TO measurements is associated with a rigid dislocation network. Ultrasonic measurement have found dislocation line densities in bulk solid $^4$He as large as $6\times10^9$ per cm$^2$.[29] Assuming that the primary silica particles form fractal aggregates with a strand diameter of ~100 Å, the density of silica strands in our (95% porous) aerogel sample is on the order of $10^{11}$cm$^{-2}$, higher than the dislocation density. It should be noted that the aerogel strands do not necessarily supplant or enhance the connectivity of the dislocation network and do not necessarily lead to a larger NCRIf. On the other hand, it seems highly probable that they provide effective pinning sites for dislocation lines.

In conclusion growing solid helium in aerogel produces highly polycrystalline samples. Although at least 1% of the solid resides on grain boundaries, the NCRIf that is observed is similar to that in high quality crystals that contain few, if any, grain boundaries. This substantially rules out any explanation of the observed phenomena in terms of flow or slippage along grain boundaries. Furthermore, the aerogel network should provide abundant pinning sites for dislocation lines. Day and Beamish attribute the destruction of supersolidity to the depinning of the dislocation network when $^3$He evaporates off the dislocation lines.[15] Within this model effective pinning of the dislocation network should lead to an enhancement of the onset temperature as compared to bulk solid. This is not observed.

This work was supported through NSF DMR-0706339 (MHWC) and DE-FG01-05ER05-02 (CAB). Use of the Advanced Photon Source at Argonne National Laboratory was supported by the U. S. Department of Energy, Office of Science, Office of Basic Energy Sciences, under Contract No. DE-AC02-06CH11357.

1. Kim, E. and M.H.W. Chan, Science, **305**, 1941 (2004).
2. Kim, E. and M.H.W. Chan, Nature, **427**, 225 (2004).
3. Rittner, A.S.C. and J.D. Reppy, Phys. Rev. Lett., **97**, 165301 (2006).
4. Rittner, A.S.C. and J.D. Reppy, Phys. Rev. Lett., **98**, 175302 (2007).
5. Aoki, Y., J.C. Graves, and H. Kojima, Phys. Rev. Lett., **99**, 015301(2007).
6. Kondo, M., S. Takada, Y. Shibayama, and K. Shirahama,.Low Temp. Phys., **148**, 695 (2007).
7. Penzev, A., Y. Yasuta, and M. Kubota, J. Low Temp. Phys., **148**, 677 (2007).
8. Clark, A.C., J.T. West, M.H.W. Chan, Phys. Rev. Lett., **99**, 135302 (2007).
9. Boninsegni, M., N. Prokof'ev, and B. Svistunov, Phys. Rev. Lett., **96**, 105301 (2006).
10. Clark, B.K. and D.M. Ceperley, Phys. Rev. Lett., **96**, 105302 (2006).
11. Boninsegni, M., A.B. Kuklov, L. Pollet, N.V. Prokof'ev, B.V. Svistunov, and M. Troyer, Phys. Rev. Lett., **97**, 080401 (2006).
12. Boninsegni, M., A.B. Kuklov, L. Pollet, N.V. Prokof'ev, B.V. Svistunov, and M. Troyer, Phys. Rev. Lett., **99**, 035301 (2007).
13. Toner, J., Phys. Rev. Lett., **1**, (2008).
14. Pollet, L., M. Boninsegni, A.B. Kuklov, N.V. Prokof'ev, B.V. Svistunov, and M. Troyer, Phys. Rev. Lett., **98**, 135301 (2007).
15. Day, J. and J. Beamish, Nature, **450**, 853 (2007).
16. Kim, E., J. S. Xia, J. T. West, X. Lin, A. C. Clark, M. H. W. Chan, arXiv:0710.3370v1, (2007).
17. Clark, A.C., J.D. Maynard, and M.H.W. Chan, Phys. Rev. B, **77**, (2008).
18. Anderson, P.W., Nature Physics, **3**, 160 (2007).
19. Wallacher, D., M. Rheinstaedter, T. Hansen, and K. Knorr, J. Low Temp. Phys., **138**, 1013 (2005).
20. Pierre, A.C. and G.M. Pajonk, Chemical Reviews, **102**, 4243 (2002).
21. Chan, M., N. Mulders, and J. Reppy, Physics Today, **49**, 30 (1996).
22. Halperin, W.P., H. Choi, J.P. Davis, and J. Pollanen, arXiv:0807:2279, (2008).
23. Porto, J.V. and J.M. Parpia, Phys. Rev. B, **59**, 14583 (1999).
24. Haard, T.M., G. Gervais, R. Nomura, and W.P. Halperin, Physica B, **284**, 289 (2000).
25. Beamish, J.R. and J.P. Franck, Phys. Rev. B, **26**, 6104 (1982).
26. Khairallah, S.A. and D.M. Ceperley, Phys. Rev. Lett., **95**, (2005).
27. Golov, A.I., I.B. Berkutov, S. Babuin, and D.J. Cousins, Physica B-Condensed Matter, **329**, 258 (2003).
28. Crowell, P.A., F.W. VanKeuls, and J.D. Reppy, Phys. Rev. B, **55**, 12620 (1997).
29. Hiki, Y. and F. Tsukuba, Phys. Rev. B, **27**, 696 (1983).